\documentclass[12pt]{article}
\usepackage{amsfonts,amstext,epsfig,amsmath}

\let\a=\alpha    
  \let\q=\theta \let\k=\kappa
     
\let\s=\sigma \let\t=\tau

\let\pa=\partial
\def\be{\begin{equation}}
\def\ee{\end{equation}}
\def\ba{\begin{array}}
\def\ea{\end{array}}

\def\dalemb#1#2{{\vbox{\hrule height .#2pt
        \hbox{\vrule width.#2pt height#1pt \kern#1pt
                \vrule width.#2pt}
        \hrule height.#2pt}}}

\newcommand{\bea}{\begin{eqnarray}}
\newcommand{\eea}{\end{eqnarray}}

\newcommand{\Tr}{{\rm Tr} }

\def\bR{{{\Bbb R}}}
\def\bC{{{\Bbb C}}}

\def\bZ{{{\Bbb Z}}}

\newcommand{\preprint}[1]{\begin{table}[t]  
             \begin{flushright}               
             {#1}                             
             \end{flushright}                 
             \end{table}}                     
\renewcommand{\title}[1]{\vbox{\center\LARGE{#1}}\vspace{5mm}}
\renewcommand{\author}[1]{\vbox{\center#1}\vspace{5mm}}
\newcommand{\address}[1]{\vbox{\center\em#1}}
\newcommand{\email}[1]{\vbox{\center\tt#1}\vspace{5mm}}

\newtheorem{thm}{Theorem}

\begin{document}
\begin{titlepage}
\preprint{hep-th/0606178 \\
DAMTP-2006-51}

\title{Two universal results for Wilson loops at strong coupling}

\author{Sean A. Hartnoll}

\address{DAMTP, Centre for Mathematical Sciences,
Cambridge University\\
Wilberforce Road, Cambridge CB3 OWA, UK}
\vspace{0.2cm}

\email{s.a.hartnoll@damtp.cam.ac.uk}

\abstract{
We present results for Wilson loops in strongly coupled gauge
theories. The loops may be taken around an arbitrarily shaped
contour and in any field theory with a dual IIB geometry of the
form $M \times S^5$. No assumptions about supersymmetry are made.
The first result uses D5 branes to show how the loop in any
antisymmetric representation is computed in terms of the loop in
the fundamental representation. The second result uses D3 branes
to observe that each loop defines a rich sequence of operators
associated with minimal surfaces in $S^5$. The action of these
configurations are all computable. Both results have features
suggesting a connection with integrability.}

\end{titlepage}

\setcounter{page}{1}

\section{Introduction}

Wilson loop operators are basic nonlocal gauge invariant
operators in any Yang-Mills theory. In the AdS/CFT correspondence
\cite{ads/cft}, Wilson loops are described by similarly basic objects
in string theory. The original strong coupling dual description of
such operators was in terms of noncompact worldsheet instantons in
the bulk geometry ending on the loop at conformal infinity
\cite{Maldacena:1998im,Rey:1998ik}. There has been a recent revival of
interest in Wilson loops
in the AdS/CFT correspondence. Of particular relevance to the present
work is the appreciation that probe D3 and D5 branes are also
important for computing certain
Wilson loop observables \cite{Drukker:2005kx, Yamaguchi:2006te,
  Hartnoll:2006hr, Yamaguchi:2006tq, Gomis:2006sb, Lunin:2006xr,
  Okuyama:2006jc, Hartnoll:2006is}.

Much of the tangible progress to date in studying Wilson loop
operators in the AdS/CFT correspondence has stemmed from the fact that
certain loops are BPS operators of the ${\mathcal{N}}=4$ super
Yang-Mills (SYM) theory. The works of \cite{Erickson:2000af,
  Drukker:2000rr} have argued that half BPS circular Wilson loops are
computed exactly, up to instanton corrections \cite{Bianchi:2002gz},
to all orders in $1/N$ and $\lambda$ by a quadratic Hermitian matrix
model. This observation has allowed some of the most precise tests yet
of the AdS/CFT duality \cite{Drukker:2005kx, Yamaguchi:2006tq,
  Okuyama:2006jc, Hartnoll:2006is}.

The results presented in this letter are universal in the sense
that they apply to a very wide range of Wilson loops at strong
coupling and do not depend on supersymmetry. Firstly, they are
applicable to Wilson loops in any gauge theory with a dual IIB
geometry of the form $M
\times S^5$. The manifold $M$ can have any asymptotically
(locally) AdS five dimensional metric with constant negative Ricci
tensor. It could be $AdS_5$, but it could also be, for instance,
the five dimensional AdS-Schwarzschild black hole. Therefore all
our results apply to Wilson loops at finite temperature. Secondly,
the results are applicable to Wilson loops where the $SU(N)$
holonomy is taken around an arbitrary curve in the boundary field
theory. Such a loop will generically not preserve any
supercharges.

Our first result, in section 2, shows that the expectation value
of a Wilson loop in the rank $k$ antisymmetric representation,
computed by a D5 brane with $k$ units of worldvolume flux
\cite{Yamaguchi:2006tq,Gomis:2006sb,Hartnoll:2006is}, is given by
a universal expression in terms of the loop in the fundamental
representation, computed by a fundamental string. The relation is
naturally expressed in terms of a Hermitian matrix model. This is
surprising given that we are a long way away from BPS loops.

Section 3 is a warm up for our second result, in section 4. We
observe that for every Wilson loop with a dual string $\Sigma
\hookrightarrow M$, there is an infinite sequence of D3 brane
immersions $\Sigma \times \Upsilon \hookrightarrow M \times S^5$,
where $\Upsilon$ is a minimal surface in $S^5$. These
configurations have a rich purely algebraic description as
holomorphic curves in ${\mathbb{CP}}^4$, and their actions are all
computable.

\section{A universal result for antisymmetric loops}

Consider a general IIB background of the form $M \times S^5$. We can write
the metric as
\be\label{eq:metric}
ds^2 = R^2 \left[ds^2_M + d\q^2 + \sin^2\q d\Omega^2_{S^4} \right] \,,
\ee
where the polar angle $\theta \in [0,\pi]$. Suppose
we have an arbitrary Wilson loop in the boundary theory that
preserves $SO(5) \subset SO(6)_R$, that is to say, a loop in
$\partial M$ times a point in $S^5$. It is well known that if the
loop is in the fundamental representation, then the state dual to
this operator is a fundamental string embedded on a minimal
surface $\Sigma \subset M$ that ends on the loop in $\partial M$.
We now derive a universal expression for a Wilson loop on the same
curve on the boundary but in the rank $k$ antisymmetric
representation.

Wilson loops in the $k$-th antisymmetric representation are dual
to a D5 brane embedded in the dual geometry carrying $k$ units of
worldvolume flux \cite{Yamaguchi:2006tq,Gomis:2006sb,Hartnoll:2006is}.
Recall that the Euclidean D5 brane action is
\be
S_{D5} = T_{D5} \int d\t d^5\sigma \sqrt{\det \left(
^\star g + 2 \pi \a' F \right)} - i g_s T_{D5} \int 2 \pi \a' F
\wedge {}^\star C_4 \,,
\ee
where $T_{D5} = N \sqrt{\lambda}/8 \pi^4 R^6$. Here and elsewhere
we freely use the relation $R^4 = \lambda \a'^2$. The relevant
part of the background four form potential is
\be
C_4 = \frac{R^4}{g_s} \left[\frac{3 (\q-\pi)}{2} - \sin^3\q \cos\q
       - \frac{3}{2} \cos\q \sin\q \right] \text{vol} S^4 \,,
\ee
where $\q$ is a polar angle in the $S^5$, as in (\ref{eq:metric}).

The D5 brane embedding we are looking for will
be $\Sigma \times S^4 \hookrightarrow M \times S^5$. The $S^4$
is set to be at an angle $\theta$ in the $S^5$. This
embedding preserves an $SO(5)$ subgroup of the R symmetry, as did
the original worldsheet $\Sigma$.
Furthermore, in order to induce a string charge, we need to have
an electric worldvolume gauge field. This field will be imaginary
because we need to set $\delta S/\delta F_{\t\sigma} = i k$ to
carry the charge of $k \in \bZ$ fundamental strings. Thus we set
\be\label{eq:defineF}
F_{\t\sigma} \equiv i F \frac{\sqrt{\lambda}}{2
\pi} \,.
\ee
In these expressions we are using $\t,\sigma$ as coordinates on the
surface $\Sigma$. The action for the D5 brane on this ansatz becomes
\be\label{eq:generalD5}
S_{D5} = \frac{N \sqrt{\lambda}}{3 \pi^2} \int d\tau d\sigma
\left[ \sin^4\q \sqrt{\det {}^* g_\Sigma - F^2} - D(\q) F \right]
\,,
\ee
where we have introduced
\be\label{eq:Dtheta}
D(\q) = \sin^3\q \cos\q + \frac{3}{2} \cos\q \sin\q -
\frac{3 (\q-\pi)}{2} \,.
\ee
In the action (\ref{eq:generalD5}), ${}^* g_\Sigma$ denotes the
metric pulled back from $M\times S^5$, with all factors of $R$
stripped off, onto the worldsheet $\Sigma$.

The equations of motion for $\theta$ and $F$ allow for constant
solutions $\theta = \theta_0$, provided that $\theta_0$ satisfies
\be\label{eq:thetaconstant}
\pi \left(\frac{k}{N} - 1\right) = \sin\q_0 \cos\q_0 - \q_0 \,.
\ee
For these solutions,
\be\label{eq:generalF}
F = - \cos\theta_0 \sqrt{\det {}^* g_\Sigma} \,.
\ee
Thus as a two form, $F$ is proportional to the volume form on
$\Sigma$. Using the constancy of $\theta$ and the above expression
for $F$, it is straightforward to show that the equations of
motion for the embedding $X^{\mu}(\sigma_a \equiv \{\tau,\sigma\})$
into $M$ become
\be\label{eq:F1eqns}
\frac{\pa \sqrt{\det {}^* g_\Sigma}}{\pa X^{\mu}} =
\pa_a \frac{\pa \sqrt{\det {}^* g_\Sigma}}{\pa \pa_a X^{\mu}} \,.
\ee
Note that $\theta$ being constant implies that ${}^* g_\Sigma$ only
depends on the embedding into $M$ and not on $S^5$.
The equations (\ref{eq:F1eqns}) are of course precisely the equations
of motion following from the Nambu-Goto string action
\be\label{eq:nambu}
S_{F1} = \frac{1}{2\pi \a'} \int d\t d\s \sqrt{\det {}^\star g_{\Sigma}} \,.
\ee
Therefore, we have found that
any embedding $\Sigma$ into $M$ of a fundamental string defines an
embedding $\Sigma \times S^4$ of a D5 brane into $M \times S^5$ for
every $k \in \bZ$. Now we need to compute their actions.

In evaluating the action of the D5 brane on the solution, we need
to add the following boundary terms. These may be thought of as
implementing the correct Neumann and Dirichlet boundary conditions
and infinity and simultaneously renormalising the action
\cite{Drukker:1999zq,Drukker:2005kx,Hartnoll:2006hr}. We have
\be\label{eq:boundary}
\left. S_{D5} \right|_{\text{bdy.}} = - \int_{\partial \Sigma} d\tau
X^{\mu} \frac{\delta S_{D5}}{\delta \partial_\sigma X^{\mu}} -
\int_{\partial \Sigma} d\tau \q
\frac{\delta S_{D5}}{\delta \partial_\sigma \q}
+ \frac{\sqrt{\lambda}}{2 \pi} \int_\Sigma d\tau d\sigma k F
\,.
\ee
In this expression we have taken $\tau$ to be the worldvolume
coordinate tangent to the boundary and $\sigma$ the worldvolume
coordinate normal to the boundary. The first of these terms is the
same as that required to renormalise the fundamental string
action. The second in fact vanishes on the solution. Using
(\ref{eq:generalF}) and constancy of $\theta$ we can evaluate the
total action to obtain
\be\label{eq:amazing}
\boxed{\left. S_{D5} \right|_{\text{renor.}} = \frac{2N}{3\pi} \sin^3 \q_0
\left. S_{F1} \right|_{\text{renor.}} \,.}
\ee
Therefore we find that for a very wide class of Wilson loops,
independently of the five dimensional background and of the shape
of the loop, the loop in the $k$-th antisymmetric representation
is given in terms of the loop in the fundamental representation
through the formula (\ref{eq:amazing}), with $\theta_0$ given by
(\ref{eq:thetaconstant}). This implies, for example, that
antisymmetric loops have exactly the same dependence on $k$ at zero
and finite temperature.

The existence of probe D5 brane embeddings at the angles
(\ref{eq:thetaconstant}) determined by worldvolume flux has been
observed in various different contexts, see for instance
\cite{Callan:1998iq,Camino:1999xx,Gomis:2002xg,Hartnoll:2006hr,
  Yamaguchi:2006tq}. The expression (\ref{eq:amazing}) is a substantial
generalisation of the known results, albeit technically
straightforward. An especially striking aspect of (\ref{eq:amazing})
is that it is the result one obtains for the antisymmetric loop in a
quadratic Hermitian matrix model \cite{Hartnoll:2006is}. That is to
say, (\ref{eq:amazing}) can be rewritten as the $N \to \infty$,
$S_{F1} \gg 1$ limit of
\be\label{eq:mm}
e^{- S_{D5,k}} = \int [dM] \Tr_{A_k}[e^{M}] e^{- 2
  N/S_{F1}^2 \Tr[M^2] } \,.
\ee
In this last expression both the D5 and F1 actions are understood to
be renormalised. The $A_k$ index indicates that the trace is taken in
the rank $k$ totally antisymmetric representation of $SU(N)$. The matrix $M$ is
an $N \times N$ Hermitian matrix.

For the circular BPS Wilson loop, $S_{F1} = - \sqrt{\lambda}$ and one
recovers the expected matrix model of \cite{Erickson:2000af,
  Drukker:2000rr}. The appearance of a matrix model in a much more
general setting clearly raises many interesting questions. We should
emphasise, however, that we have only made the connection in the large
$N$, large $\lambda$ limit. There may be many matrix model and non
matrix model computations other than (\ref{eq:mm}) that lead to the
same answer (\ref{eq:amazing}) in that limit.

\section{Some D3 brane embeddings}

One might search for a similar universal result for Wilson loops
in the rank $k$ symmetric representation. These are probably dual
to a D3 brane in the bulk geometry
\cite{Gomis:2006sb,Hartnoll:2006is}. This appears to be a harder
question than the antisymmetric case and we will not address it
here. Instead, we describe some intriguing D3 brane solutions that
are related to Wilson loops.

The reason that the symmetric representations are difficult is the following.
We expect the worldvolume flux and background Ramond-Ramond potential to
blow up the worldsheet from $\Sigma$ to $\Sigma \times S^2$. However,
in order to preserve the $SO(5)$ R-symmetry of the Wilson loop, the
whole $\Sigma \times S^2$ worldvolume should be embedded into $M$
rather than $M \times S^5$. The required D3 brane solution has been
found for the case of a circular loop when $M = AdS_5$
\cite{Drukker:2005kx}. However, for
more general loops, or more general backgrounds with less or no symmetry,
finding the embedding appears to be a daunting task that needs to
be approached on a case by case basis.

In this section and the next we will look for D3 brane solutions which
are similar to the D5 brane solutions of the previous section. They
will be the product of a string embedding $\Sigma$ into $M$ and a map
from $S^2$ into $S^5$. To begin with we require $S^2$ to be embedded
into $S^5$. We then relax this restriction to allow
self intersections, revealing a rich mathematical structure underpinning the
solutions.

Let us write the background metric as follows
\be
\label{eq:metric2}
ds^2 = R^2 \left[ds^2_M + d\a^2 + \sin^2\a d\Omega^2_{S^2} + \cos^2\a
  d\widetilde \Omega^2_{S^2} \right] \,,
\ee
where $\a \in [0,\pi/2]$. Given an arbitrary fundamental string embedding of
$\Sigma$ into $M$, we will show in this section that there is a D3 embedding of
$\Sigma \times S^2 \hookrightarrow M \times S^5$ carrying $k$ units of
worldvolume flux. This embedding will break the $SO(5)$ symmetry of
$\Sigma$ down to $SO(3) \times SO(2)$.

The relevant D3 brane action is
\be\label{eq:d3action}
S_{D3} = T_{D3} \int d\t d^3\sigma \sqrt{\det \left( ^\star g
+ 2 \pi \a' F \right)}\,,
\ee
where the tension $T_{D3} = N/2\pi^2 R^4$. We have not written down
the Wess-Zumino term as this will not contribute to the configurations
under consideration, which have two directions in $M$ and two
directions in $S^5$. More precisely, the embedding we look for will
wrap the first $S^2$ in (\ref{eq:metric2}) at some value of
$\a$. As before, we set
\be
F_{\t \sigma} \equiv i F \frac{\sqrt{\lambda}}{2 \pi} \,.
\ee
The action for the D3 brane becomes
\be
S_{D3} = \frac{2 N}{\pi} \int d\t d\sigma \sin^2\a \sqrt{\det {}^*
  g_\Sigma - F^2} \,.
\ee
Here again, ${}^* g_\Sigma$ denotes the metric pulled back from the
embedding onto $\Sigma$.

The equation of motion for $\alpha$ allows for two constant solutions,
$\a = \a_0 = 0$ and $\a = \a_0 = \pi/2$. The first of these
corresponds to the $S^2$ collapsing to a point and takes us outside
the regime of validity of the Dirac-Born-Infeld action
(\ref{eq:d3action}). The second is fully reliable and non
collapsed. On these solutions, the equation of motion for $F$ may be
solved to give
\be
F = \frac{\k}{\sqrt{\sin^4\a_0 + \k^2}} \sqrt{\det {}^* g_\Sigma} \,.
\ee
Once again, the worldvolume gauge field is proportional to the volume
form on the worldsheet. As is usual in D3 brane computations
\cite{Drukker:2005kx,Hartnoll:2006hr}, we have introduced
\be
\kappa = \frac{\sqrt{\lambda} k}{4 N} \,.
\ee
It is then immediate to check that the embedding of $\Sigma$ into
$M$ must satisfy the equations following from the Nambu-Goto
action (\ref{eq:nambu}). Therefore, for any embedding of a
fundamental string into $M$ we have found an embedding of a probe
D3 brane into $M
\times S^5$ for all $k \in \bZ$.

To renormalise the action and impose the correct boundary conditions
at infinity, we need to add the boundary term
\be
\left. S_{D3} \right|_{\text{bdy.}} = - \int_{\partial \Sigma} d\tau
X^{\mu} \frac{\delta S_{D3}}{\delta \partial_\sigma X^{\mu}} -
\int_{\partial \Sigma} d\tau \a
\frac{\delta S_{D3}}{\delta \partial_\sigma \a}
+ \frac{2 N}{\pi} \int_\Sigma d\tau d\sigma \k F
\,.
\ee
The effects of these terms are similar to what occurred in the D5
brane case. The renormalised action becomes
\be\label{eq:nice}
\boxed{\left. S_{D3} \right|_{\text{renor.}} = \frac{4 N}{\sqrt{\lambda}}
\sqrt{\sin^4\a_0 + \k^2} \left. S_{F1} \right|_{\text{renor.}} \,.}
\ee
If the renormalised F1 action is negative, as it is for the
cases that have been computed, then the blown up solution $\a_0 =
\pi/2$ is favoured over the collapsed solution $\a_0 = 0$.
Assuming negativity of the F1 action, (\ref{eq:nice}) may be
suggestively rewritten as
\be
S_{D3,k} = - \sqrt{\left( k S_{F1} \right)^2 + \left(
  S_{D3,0}\right)^2} \,,
\ee
with all actions in this expression understood to be renormalised.

We can see from (\ref{eq:nice}) that as $k/N \to 0$ the action
does not tend to that of $k$ fundamental strings, suggesting that
these solutions should not be understood as blown up fundamental
strings. Indeed, the `blown up' solution remains favoured over the
collapsed solution even when $k = 0$. Independently of $k$, and
because $F$ is proportional to the volume form on $\Sigma$, the D3
branes are simply minimal surfaces in $M \times S^5$. Such brane
configurations usually have negative modes due to the fact that
the $S^2$ is contractible within $S^5$. However, the renormalised
area of $\Sigma$ in $M$ is negative for all known solutions when
$M$ is $AdS_5$ or Schwarzschild-$AdS_5$. Therefore minimising the
action (\ref{eq:nice}) favours larger areas of $S^2
\subset S^5$ and the collapsing modes increase rather than
decrease the action. This counterintuitive behaviour seems to be
related to the imposition of Neumann rather than Dirichlet
boundary conditions at infinity for certain of the embedding
fields.

\section{D3 brane immersions and minimal surfaces in $S^5$}

It becomes clear at this point that we need not have restricted
our D3 brane configurations to such a symmetric ansatz. Consider any
F1 embedding $\Sigma \hookrightarrow M$, dual to a fundamental
representation Wilson loop in the gauge theory. Given any minimal two dimensional
surface $\Upsilon$ in $S^5$ then $\Sigma \times \Upsilon
\hookrightarrow M \times S^5$ gives a
solution to the D3 brane equations of motion, again for any $k \in
\bZ$. A straightforward adaptation of the argument in the previous
section shows that the action of this configuration is
\be\label{eq:minimal}
\boxed{\left. S_{D3} \right|_{\text{renor.}} = \frac{4 N}{\sqrt{\lambda}}
\sqrt{\left(\text{Vol} \Upsilon/4\pi\right)^2 + \k^2} \left. S_{F1}
\right|_{\text{renor.}} \,.}
\ee

As well as the high symmetry, a special feature of the solution of the
previous section is that it was an embedding of $S^2$ into $S^5$.
Generic minimal surfaces are not embeddings, but rather they are
immersions. That is to say, they self intersect. If we allow the
D3 brane to self intersect, then the solution we described in the
previous section is just the tip of an iceberg. There does
not appear to be any reason to discard self intersecting
immersions. It is likely that $\a'$ corrections will disconnect the
brane at the intersection by condensing a tachyonic mode, but at a
classical level the self intersecting solution provides an adequate
description.

Let us restrict ourselves to the case that $\Upsilon$ is a minimal
immersion of $S^2$ into $S^5$. Higher genus immersions are
certainly interesting, but we wish to make use here of the highly
developed mathematical theory of immersions of spheres into
spheres. The following theorems are applicable to such immersions.
A review of this material and some additional theorems can be
found, for instance, in \cite{review}.

\begin{thm} {\bf (Calabi \cite{Calabi}, Barbosa \cite{Barbosa})}
  The minimal immersion lies in
  an extremal $S^4 \subset S^5$. Either it is a totally geodesic $S^2$ with
  area $4\pi$ or it has area $4\pi m$, with $m \in \bZ$ and $m \geq 3$.
\end{thm}

The totally geodesic $S^2$s are the cases that we considered in the
previous section. This theorem implies that we can set $\text{Vol}
\Upsilon/4\pi = m$ in (\ref{eq:minimal}). Thus the actions of these
configurations are all given in terms of the fundamental string
action and an integer, without needing to know the precise form of
$\Upsilon$. Before moving on to a general description, note that
the cases in which the induced metric on $\Upsilon$ has constant
curvature can be characterised very explicitly:

\begin{thm} {\bf (Calabi \cite{Calabi})} If the induced metric on
  $\Upsilon$ has constant curvature, then $\Upsilon$ is either a
  totally geodesic $S^2$ or it is the Veronese surface.
\end{thm}

The Veronese surface in this context is the immersion of $\bR^3$ into $\bR^6$
\be\label{eq:veronese}
\left(x_1,x_2,x_3\right) \to
\left(\sqrt{3}x_1 x_2,\sqrt{3}x_1 x_3,\sqrt{3} x_2 x_3,{\textstyle \frac{\sqrt{3}}{2}} (x_1^2-x_2^2),{\textstyle
  \frac{1}{2}} (x_1^2+x_2^2-2x_3^2),0\right) \,.
\ee
When restricted to $x^2_1+x_2^2+x_3^2 = 1$, this gives a minimally
immersed $S^2$ in $S^4 \subset S^5$ with constant curvature and
area $12 \pi$, corresponding to $m=3$ in Theorem 1. Of course,
this surface and all immersions are defined up to rigid $SO(6)$
rotations on $S^5$.

We now describe general minimal immersions. From Theorem 1 it is
sufficient to consider maps $x:S^2 \to S^4$. We will describe $x$ as
a unit vector in $\bR^5$. We can always choose
isothermal coordinates $\{z,\bar z\}$ on $S^2$ so that the induced metric is
\be\label{eq:induced}
ds^2_{\Upsilon} = 2 \pa x \cdot \bar \pa x \, dz d\bar z \,.
\ee
Here and below $a \cdot b$ denotes the symmetric dot product on
$\bC^5$, inherited from $\bR^5$. With these coordinates, minimal
immersions are those that satisfy the nonlinear equation
\be\label{eq:nonlinear}
\pa \bar \pa x = - \pa x \cdot \bar \pa x \, x \,.
\ee
Remarkably, this equation can be solved in full generality for
immersions of $S^2$ into $S^4$.

\begin{thm} {\bf (Chern \cite{Chern}, Barbosa \cite{Barbosa})} There
  is a canonical one to one correspondence between minimal immersions
  $x: S^2 \to S^4$ and the set of totally isotropic holomorphic curves
  $\xi: S^2 \to {\mathbb{CP}}^4$.
\end{thm}

The general minimal immersion is constructed as follows. Let $\xi$
be homogeneous coordinates on ${\mathbb{CP}}^4$. The holomorphic
curves are necessarily polynomials, so that
\be
\xi(z) = \sum_{i=0}^n A_i z^i \,,
\ee
with $A_i \in \bC^5$. Total isotropy is the statement that the
polynomial must satisfy
\be
\xi \cdot \xi = \pa \xi \cdot \pa \xi = 0 \,.
\ee
Given such a polynomial we define, using roman indices running
from 1 to 5,
\be\label{eq:psi}
\psi_a = \epsilon_{a b c d e} \, \xi_b \, \pa \xi_c \, \bar \xi_d \, \bar
\pa \bar \xi_e \,.
\ee
It is clear that $\psi \in \bR^5$. The minimal immersion is then given by
\be\label{eq:x}
x = \frac{\psi}{\mid \psi \mid} \,.
\ee

For example, the following curve in ${\mathbb{CP}}^4$ describes
the Veronese immersion (\ref{eq:veronese})
\bea
& \xi_1 = 2 \sqrt{3} (1+z^4) \,, \quad \xi_2 = 2 \sqrt{3} i
(1-z^4) \,, \quad \xi_3 = -12 z^2 \,, \nonumber \\
& \xi_4 = - 4 \sqrt{3} i (z+z^3) \,, \quad \xi_5 =  - 4 \sqrt{3}
(z-z^3)\,.
\eea
Substituting this expression into (\ref{eq:psi}) and (\ref{eq:x})
and changing the coordinates on $S^2$ to $z = (x_1+i x_2)/(1+x_3)$
with $x^2_1+x_2^2+x_3^2 = 1$, one recovers the immersion
(\ref{eq:veronese}). Using this formalism, Barbosa has constructed
examples of regular immersions with all values of the area $4\pi
m$ allowed by Theorem 1 \cite{Barbosa}. That paper also contains
other interesting results on the moduli space of immersions and
their regularity. Minimal immersions in $S^4$ can also be described
using the Penrose twistor transform. In that description they are
given by holomorphic curves in ${\mathbb{CP}}^3$ that are horizontal
with respect to the twistor fibration over $S^4$ \cite{bryant}.

The point we wish to emphasise following from the above discussion
is this: to every Wilson loop operator of the boundary theory,
with dual F1 embedding $\Sigma \hookrightarrow M$, we can
associate an infinite sequence of distinct Wilson loop operators,
one for each minimal surface $\Upsilon \hookrightarrow S^5$. These
new operators generically preserve no R symmetry. Nonetheless, the
action of the dual D3 solutions is determined entirely in terms of
the original F1 action and a single integer $m$. A priori, each
operator is associated with a solution to a nonlinear partial
differential equation. However, we have seen that in fact they are
counted by the purely algebraic problem of finding totally
isotropic holomorphic curves on ${\mathbb{CP}}^4$.

What are these additional operators? We will not attempt to answer
this question here. Note that
Wilson loop operators can be
obtained by integrating any adjoint valued operator around a closed
contour $C$ in spacetime. Using only the six adjoint scalars $\{\Phi_I\}$ of
the ${\mathcal{N}}=4$ SYM theory, a natural set of operators to
consider are
\be\label{eq:operator}
W[C,\Upsilon] = \Tr \, {\mathcal{P}} \exp \left[ i \oint_C ds \left(
  A(s) + i \Theta_I \Phi_I(s) + \sum_{n \geq 2} f_{I_1 \cdots I_n}
  \Phi_{I_1}(s) \cdots \Phi_{I_n}(s) \right) \right] \,,
\ee
with coefficients $f_{I_1 \cdots I_n}$ defined in terms of
integrals over the immersion $\Upsilon$. In this expression $A$ is
the $SU(N)$ gauge field. The first two terms in the exponent are
those appearing in the usual Euclidean Maldacena-Wilson loop of
the ${\mathcal{N}}=4$ theory. For the loops we have been
considering, with $\Sigma \hookrightarrow M$, $\Theta^I$ is a
constant unit vector in $\bR^6$. A simple possibility for the
first coefficient beyond the standard term is
\be
f_{I J} = \int_{\Upsilon} \sqrt{\det {}^* g_{\Upsilon}} {}^*
g_{\Upsilon}^{ij} \, \pa_i x_I
\pa_j x_J = \int_{\Upsilon} dz d\bar z \, \pa x_I \bar \pa x_J \,,
\ee
where $x_I(z,\bar z)$ is the immersion of $\Upsilon$ into $S^5 \subset
\bR^6$ and ${}^* g_{\Upsilon}$ is the induced metric (\ref{eq:induced}).
The definition of these operators should not require $\Upsilon$ to
be a minimal surface. When the surface is minimal, however, we
have seen that the operators have a particularly simple dual
description. The elegant description of this class of Wilson loop
operators at strong coupling clearly warrants a field theory
understanding.

\section{Concluding remarks}

We have not yet pursued a field theory understanding of the
results in this letter. The possible
appearance of a matrix model in the
study of antisymmetric representations, and of an integrable
nonlinear partial differential equation underpinning the D3 brane
configurations, suggests that perhaps similar integrable
structures should be present in the field theory duals.

For the case of the D3 brane configurations, it is important to
understand whether there are normalisable negative modes about the
solutions. The minimal immersions $\Upsilon \hookrightarrow S^5$
define the set of possible asymptotic configurations near the
boundary. However, if there is a negative mode then the dominant
contribution to the expectation value will come from a more
complicated configuration than $\Sigma \times \Upsilon$. It would
also be interesting to understand the case when $\Upsilon$ is of
higher genus.

\section*{Acknowledgements}

I would like to thank Gary Gibbons, Nick Dorey and especially Prem
Kumar for helpful comments. The author is supported by a research fellowship
from Clare college, Cambridge.

\end{document}